\documentclass[reqno,10pt,a4paper]{amsart}

\usepackage{amssymb}
\usepackage{url}
\usepackage{hyperref}
\allowdisplaybreaks[4]

\usepackage{mathrsfs}
\let\mathcal\mathscr
\usepackage[mathscr]{eucal}

\usepackage{color}

\let\phi=\varphi
\let\kappa=\varkappa
\DeclareMathOperator{\rank}{rank}

\newcommand*{\eval}[1]{\left.#1\right|}
\newcommand*{\abs}[1]{\left|#1\right|}

\theoremstyle{theorem}

\theoremstyle{definition}

\theoremstyle{remark}
\newtheorem{remark}{Remark}

\usepackage[all]{xy}
\SelectTips{cm}{}

\usepackage{mathrsfs}
\let\mathcal\mathscr
\usepackage[mathscr]{eucal}

\newcommand{\cprime}{\/{\mathsurround=0pt$'$}}

\begin{document}
\title[RO for the $2$-component rdDym]{Recursion operators in the cotangent
  covering of the rdDym equation}

\author{I.S.~Krasil{\cprime}shchik} \address{Trapeznikov Institute of Control
  Sciences, 65 Profsoyuznaya street, Moscow 117997, Russia}
\email{josephkra@gmail.com}\thanks{ The work ISK was partially supported by
  the RFBR Grant 18-29-10013 and RSF Grant 21-71-20034.} \author{A.M.~Verbovetsky}
\address{Independent University of Moscow, Bolshoy Vlasyevskiy
  Pereulok 11, Moscow, 119002, Russia} \email{alik@ejik.org}

\begin{abstract}
  We describe a general method of constructing nonlocal recursion operators
  for symmetries of PDEs. As an example, the cotangent equation to the 3D
  rdDym equation $u_{yt} = u_xu_{xy} - u_yu_{xx}$ for which two mutually
  inverse operators are found. The exposition includes a rigorous criterion to
  check the hereditary property.
\end{abstract}

\keywords{Partial differential equations, integrable linearly degenerate
  equations, nonlocal symmetries, recursion operators}

\subjclass[2010]{35B06}

\maketitle

\tableofcontents

\section*{Introduction}
\label{sec:introduction}
The rdDym equation ($r$th dispersionless Dym equation)
\begin{equation}
  \label{eq:1}
  u_{yt} = u_xu_{xy} - u_yu_{xx}
\end{equation}
belongs to the class of linearly degenerate equations~\cite{Fer-Mos} and
arose in the Mart\'{\i}nez Alonso-Shabat hierarchy~\cite{MA-Sh}. It is
Lax integrable and its symmetry properties were studied
in~\cite{BKMV-2016, Baran-etal-compar-stud}. Recursion operators for
symmetries of Equation~\eqref{eq:1} was found by O.~Morozov
in~\cite{Morozov2012}.

In~\cite{Ovsienko2010}, V.~Ovsienko considered (in slightly different
notation) the system
\begin{equation}
  \label{eq:2}
  \begin{array}{rcl}
    v_{yt} &=& 2 (v_yu_{xx} - v_xu_{xy}) + u_xv_{xy} - u_yv_{xx}
               -2 (u_yu_{xx} + 2 u_xu_{xy}),\\[2pt]
    u_{yt} &=& u_xu_{xy} - u_yu_{xx},
  \end{array}
\end{equation}
which is a two-component extension of the rdDym equation (another
two-component generalization of~\eqref{eq:1} was studied
in~\cite{Morozov2012_SIGMA}). Treating~\eqref{eq:2} as the Euler equation on
the Virasoro algebra, its bi-Hamiltonian structure was established.

System~\eqref{eq:2} may be also considered in the following way. Recall
(see~\cite{ell-star, Kra-Ver-Vit-Springer}) that to any equation
$\mathcal{E} = \{F[u] = 0\}$ in unknowns~$u$ its cotangent
equation~$\mathcal{T}^*\mathcal{E}$
\begin{equation*}
  F[u] = 0, \qquad \ell_F^*[u, p] = 0
\end{equation*}
is naturally associated, $\ell_F^*$ is the operator adjoint to the
linearisation of~$F$ (here and below $[u]$, $[u, p]$, etc. denotes the
variables together with their derivatives up to certain
order). Then~\eqref{eq:2} may be understood as the cotangent equation
of~\eqref{eq:1}, with~$v$ instead of~$p$, `deformed' by the Lagrangian term
\begin{equation*}
   \frac{\delta L}{\delta u}\ \text{ with }\ L = u_yu_x^2.
\end{equation*}
System~\eqref{eq:2} is an Euler-Lagrange equations.

Lagrangian deformations of some linearly degenerate equations were studied
in~\cite{Baran-etal-Lagr-deform}. In particular, the system
\begin{equation}
  \label{eq:3}
  \begin{array}{rcl}
    v_{yt} &=& 2 (v_y u_{xx} - v_x u_{xy})+ u_x v_{xy} - u_y
               v_{xx} - 2(2 k 
               u_{xy} u_x + (k u_y + l) u_{xx} ),\\[2pt]
    u_{yt} &=& u_xu_{xy} - u_yu_{xx},
  \end{array}
\end{equation}
which is considered below, is the deformation of the cotangent to rdDym
equation with the deforming Lagrangian $L = (ku_y + l)u_x^2$, $k$,
$l\in\mathbb{R}$. System~\eqref{eq:2} is obtained from~\eqref{eq:3}
when~$k = 1$, $l = 0$. We find below recursion operators for symmetries
of~\eqref{eq:3}.

\begin{remark}[courtesy O.I.~Morozov]
  Actually, both parameters~$k$ and~$l$ are fake in a sense: the change of
  variables $v \mapsto v - 2kxu_x - ku + ly$ kills them and transforms
  System~\eqref{eq:3} to the pure cotangent equation of~\eqref{eq:1}. In
  particular, the Ovsienko system~\eqref{eq:2} also reduces to the cotangent
  equation. Though the computations in the general case are not much harder
  that in a particular one, the result look more complicated. For that reason
  we set~$k = l = 0$ everywhere below, i.e., the equation we study below will
  be of the form
  \begin{equation}
    \label{eq:15}
    \begin{array}{rcl}
      v_{yt} &=& 2 (v_y u_{xx} - v_x u_{xy})+ u_x v_{xy} - u_y
                 v_{xx},\\[2pt]
      u_{yt} &=& u_xu_{xy} - u_yu_{xx}.
    \end{array}
  \end{equation}
\end{remark}

There exist several approaches to construct recursion operators (see, for
example,~\cite{Morozov2014_CEJM, Sergyeyev2015}). Our method is based on the
technology described in~\cite{Kra-Ver-Vit-Springer} and originating
from~\cite{Ker-Kras-Kluwer}. It is based on geometrical interpretation of
recursion operators for symmetries as B\"{a}cklund auto-transformations of the
equation tangent to the given one (cf.~\cite{Mar-another}) and essentially
uses the theory of differential coverings~\cite{VinKrasTrends}.

\begin{remark}
  Recursion operators as B\"{a}cklund auto-transformations of the tangent
  equations initially aroused in M.~Marvan's paper~\cite{Mar-another}, though
  implicitly the tangent equation as an instrument for computation of
  recursion operators appears in~\cite{Twente-Memo}. Informally, tangent and
  cotangent bundles to PDEs are discussed in the paper~\cite{Kuper-dark-eqs}
  by B.~Kupershmidt. A detailed discussion of the tangent covering and its
  role in the theory of Hamiltonian (Poisson) structures can be found
  in~\cite{ell-star} and especially in~\cite{K-V-Gdeq}.
\end{remark}

The necessary theoretical matters needed for computations are shortly
described in Section~\ref{sec:preliminaries}, while
Section~\ref{sec:main-result} contains the main results.

\section{Preliminaries and notation}
\label{sec:preliminaries}

The exposition here is based on~\cite{AMS, Kra-Ver-Vit-Springer,
  VinKrasTrends}.

Let $\pi\colon E\to M$ be a vector bundle over a smooth manifold~$M$,
$\dim M = n$, $\rank\pi = m$, and $\pi_\infty\colon J^\infty(\pi)\to M$ be the
bundle of its infinite jets. Consider an infinitely prolonged differential
equation $\mathcal{E}\in J^\infty(\pi)$ and assume that
$\mathcal{E} = \{F = 0\}$, where $F\in \Gamma(\pi_\infty^*(\xi))$, where
$\xi\colon N\to M$ is another vector bundle. Any~$\mathcal{E}$ is naturally
endowed with an $n$-dimensional Frobenius integrable
distribution~$\mathcal{C}$ (the Cartan distribution) that defines its
geometry. The Cartan distribution is $\pi_\infty$-horizontal and,
consequently, defines a flat connection
in~$\pi_\infty\colon \mathcal{E}\to M$.

A morphism of equations $f\colon\tilde{\mathcal{E}}\to \mathcal{E}$ is a
smooth map, such that
$f_*(\tilde{\mathcal{C}}_\theta)\subset \mathcal{C}_{f(\theta)}$ for
any~$\theta\in \tilde{\mathcal{E}}$. A morphism is called a (differential)
covering if it is a submersion and the
restriction~$\eval{f_*}_{\mathcal{C}_\theta}$ of its differential to any
Cartan plane is an isomorphism. Two coverings $f_i\colon \mathcal{E}_i \to
\mathcal{E}$, $i=1$, $2$, are called equivalent if there exists a
diffeomorphism $g\colon \mathcal{E}_1 \to \mathcal{E}_2$ that preserves the
Cartan distributions and such that $f_1 = f_2\circ g$.

Denote by $\mathcal{F} = \mathcal{F}(\mathcal{E})$ the algebra of smooth
functions on~$\mathcal{E}$. A (higher infinitesimal) symmetry of~$\mathcal{E}$
is a $\pi_\infty$-vertical derivation $S\colon \mathcal{F}\to \mathcal{F}$ (a
vector field on~$\mathcal{E}$) that preserves the Cartan distribution. There
exists a one-to-one correspondence between symmetries and solutions of the
equation
\begin{equation*}
  \ell_{\mathcal{E}}(\phi) = 0,
\end{equation*}
where $\ell_{\mathcal{E}}$ is the restriction of the linearisation
$\ell_F\colon \kappa\to P$ to~$\mathcal{E}$ and~$\kappa$ denotes the module of
sections~$\Gamma(\pi_\infty^*(\pi))$. A conservation law of~$\mathcal{E}$ is a
$\pi_\infty$-horizontal form $\omega\in\Lambda_h^{n-1}(\mathcal{E})$ closed
with respect to the horizontal de~Rham differential~$d_h$. Trivial
conservation laws are those of the form~$\omega = d_h\rho$, $\rho\in
\Lambda^{n-2}(\mathcal{E})$.

Given a covering $\tau\colon \tilde{\mathcal{E}} \to \mathcal{E}$, symmetries
and conservation laws of~$\tilde{\mathcal{E}}$ are called nonlocal
for~$\mathcal{E}$. Let $S\colon \mathcal{F}(\mathcal{E}) \to
\mathcal{F}(\tilde{\mathcal{E}})$ be a $\pi_\infty$-vertical derivation. It is
called a $\tau$-shadow if
\begin{equation*}
  \tilde{\mathcal{C}}_X\circ S = S\circ \mathcal{C}_X
\end{equation*}
for any vector field $X$ on~$M$. We say that a symmetry~$\tilde{S}$ is a lift
of the shadow~$S$ if~$\eval{\tilde{S}}_{\mathcal{E}} = S$.

Let an equation~$\mathcal{E}$ be given by~$\{F = 0\}$. Then the
system~$\mathcal{T}\mathcal{E}$
\begin{equation*}
  F[u] = 0, \qquad \ell_F[u, q] =0
\end{equation*}
is called the tangent equation of~$\mathcal{E}$, while the projection
$\mathrm{t}\colon [u, q]\mapsto [u]$ is the tangent covering
over~$\mathcal{E}$. It must be stressed that~$q$ is considered as an odd
variable of degree~$1$. Sections of~$\mathrm{t}$ that preserve the Cartan
distributions are identified with symmetries of~$\mathcal{E}$ and consequently
B\"{a}cklund auto-transformations~$\mathcal{R}_{\tau,\tau'}$ of the form
\begin{equation}\label{eq:6}
  \xymatrix{
    &W\ar[dl]_-{\tau}\ar[dr]^-{\tau'}&\\
    \mathcal{T}\mathcal{E}&&\mathcal{T}\mathcal{E}\rlap{,}
  }
\end{equation}
where $\tau$, $\tau'$ are coverings, are naturally interpreted as recursion
operators for symmetries of~$\mathcal{E}$.

To construct these operators, we use a scheme consisting of two steps. In the
first step, we are looking for two-component conservation laws~$\omega$
of~$\mathcal{T}\mathcal{E}$ linear on the fibers of the
projection~$\mathrm{t}$. Provided such a law was found, we consider the
covering $\tau = \tau_\omega$ associated with it and, as the second step, try
to find $\tau_\omega$-shadows (also linear on fibers both of~$\mathrm{t}$
and~$\tau_\omega$). If such a shadow~$S$ exists, it delivers the
covering~$\tau' = \tau_S$ and the desired recursion
operator~$\mathcal{R}_{\tau_\omega,\tau_S}$.

Finally, few words about the type of equations that are the subject of our
study. Assume~$\mathcal{E} = \{F = 0\}$, $F\in P$, and consider the operator
$\ell_F^*\colon \hat{P} \to \hat{\kappa}$ formally adjoint to~$\ell_F$, where
$\hat{\bullet} = \hom(\bullet, \Lambda_h^n)$. Then the
system~$\mathcal{T}^*\mathcal{E}$
\begin{equation}
  \label{eq:4}
  \ell_F^*[u, p] = 0,\qquad F[u] = 0
\end{equation}
is said to be the cotangent equation of~$\mathcal{E}$, while the projection
$\mathrm{t}^*\colon \mathcal{T}^*\mathcal{E} \to \mathcal{E}$, $(u, p) \mapsto
(u)$, is called the cotangent covering. Similar to the case of the tangent
equation, the variable~$p$ is odd. System~\eqref{eq:4} is always an
Euler-Lagrange equation with the Lagrangian density $L = \sum p^jF^j$ and we
are interested in equations
\begin{equation*}
  \ell_F^*[u, p] + G[u, p] = 0,\qquad F[u] = 0,
\end{equation*}
(`Lagrangian deformations' of~\eqref{eq:4}), such that (a) they are Lagrangian
as well and (b) they possess a nontrivial recursion
operator. Equation~\eqref{eq:2} is exactly of this type. Note that for
Euler-Lagrange equations~$\mathcal{T}\mathcal{E}$
and~$\mathcal{T}^*\mathcal{E}$ coincide, since $\ell_F = \ell_F^*$ in this
case.

\subsection*{Coordinates}
\label{sec:coordinates}
Let $\mathcal{U}\subset M$ be a coordinate neighborhood with local coordinates
$x = (x^1,\dots,x^n)$ and $E\supset\pi^{-1}(\mathcal{U})\simeq
\mathcal{U}\times \mathbb{R}^m$ be a trivialization of~$\pi$ with fiber-wise
coordinates $u = (u^1,\dots, u^m)$. Then adapted coordinates~$u_\sigma^j$
arise on~$J^\infty(\pi)$, where~$u_\sigma^j$ correspond to the partial
derivative $\partial^{\abs{\sigma}}u^j/\partial x^\sigma$. The Cartan
distribution is spanned by the total derivatives
\begin{equation*}
  D_i = \frac{\partial}{\partial x^i} + \sum u_{\sigma
    i}^j\frac{\partial}{\partial u_\sigma^j},
\end{equation*}
while the Cartan connection takes~$\partial/\partial x^i$
to~$D_i$. Equation~$\mathcal{E}$ is given by the system of relations~$F^l(x,
\dots, u_\sigma^j,\dots) = 0$, $l = 1,\dots,r$. To restrict necessary objects
to~$\mathcal{E}$, one needs to choose internal coordinates on~$\mathcal{E}$
(cf.~\cite{Mar-ort}) and express these objects in terms of internal
coordinates. In particular, the Cartan distribution on~$\mathcal{E}$ is got by
rewriting the operators~$D_i$ in internal coordinates.

The coordinate presentation of the linearisation operator acts on $\phi =
(\phi^1, \dots, \phi^m)$ by
\begin{equation*}
  \ell_F(\phi) = \sum_{\sigma, j}\frac{\partial F^l}{\partial
    u_\sigma^j}D_\sigma(\phi^j), 
\end{equation*}
where~$D_\sigma$ is the composition of the total derivatives corresponding to
the multi-index~$\sigma$, while for the adjoint one we have
\begin{equation*}
  \ell_F^*(\psi) = \sum_{\sigma,
    l}(-1)^{\abs{\sigma}}D_\sigma\left(\frac{\partial F^l}{\partial
      u_\sigma^j}\psi^l\right), 
\end{equation*}
where~$\psi = (\psi^1, \dots, \psi^r)$. Thus, the tangent equation
of~$\mathcal{E}$ is
\begin{equation*}
  \sum_{\sigma, j}\frac{\partial F^l}{\partial
    u_\sigma^j}q_\sigma^j = 0,\quad F(x,\dots, u_\sigma^j, \dots) = 0,\qquad q
  = (q^1,\dots,q^m),
\end{equation*}
while while the cotangent one is given by
\begin{equation*}
  \sum_{\sigma,
    l}(-1)^{\abs{\sigma}}D_\sigma\left(\frac{\partial F^l}{\partial
      u_\sigma^j}p^l\right) = 0,\quad F(x,\dots, u_\sigma^j, \dots) = 0,\qquad p
  = (p^1,\dots,p^r).
\end{equation*}

Without loss of generality, we can define two-component conservation laws as
the forms
\begin{equation*}
  \omega = (X_1\,dx^1 + X_2\,dx^2)\,dx^3\wedge\dots\wedge\,dx^n, \qquad
  D_1(X_2) = D_2(X_1),
\end{equation*}
where~$X_1$ and~$X_2$ are smooth functions on~$\mathcal{E}$. Triviality
of~$\omega$ means existence of a potential~$Y$, such that~$D_i(Y) = X_i$, $i =
1$, $2$.

A covering structure in a locally trivial vector bundle $\tau\colon W \to
\mathcal{E}$ is determined by the system of pair-wise commuting vector fields
\begin{equation}\label{eq:5}
  \tilde{D}_i = D_i + X_i, \qquad i =  1, \dots,n,
\end{equation}
where $X_i = \sum X_i^\alpha\partial/\partial w^\alpha$ are $\tau$-vertical
fields and~$w^\alpha$ are coordinates in the fiber of~$\tau$ (nonlocal
variables). The equalities~$[\tilde{D}_i, \tilde{D}_j] = 0$ amount to the fact
that the system
\begin{equation*}
  w_{x^i}^\alpha = X_i^\alpha,\qquad i=1,\dots,n,\quad \alpha=1,\dots,\rank \tau,
\end{equation*}
is compatible modulo~$\mathcal{E}$. Note that~\eqref{eq:5} allows to lift any
differential operator~$\Delta$ in total derivatives from~$\mathcal{E}$ to an
operator~$\tilde{\Delta}$ on the covering equation.

Let~$\omega$ be a two-component conservation law like above. Then one can
construct the covering~$\tau_\omega$ as follows. In the case~$\dim M = 2$ we
set
\begin{equation*}
  w_{x^1} = X_1, \qquad w_{x^2} = X_2,
\end{equation*}
and $\rank\tau_\omega = 1$. When~$\dim M > 2$, we introduce infinite number of
nonlocal variables~$w^\sigma$, where~$\sigma$ is a multi-index consisting of
integers $2,\dots, n$, $\abs{\sigma}\geq 0$, and set
\begin{equation*}
  w_{x^i}^\sigma =
  \begin{cases}
    D_\sigma(X_i),&i = 1,2,\\
    w^{\sigma i},&i>2.
  \end{cases}
\end{equation*}
Obviously, we obtain covering structures in both cases.

Consider an $m$-component function $S = (S^1,\dots, S^m)$. It defines a
$\tau$-shadow if and only if
\begin{equation*}
  \tilde{\ell}_{\mathcal{E}}(S) = 0,
\end{equation*}
where~$\tilde{\ell}_{\mathcal{E}}$ is the natural lift of the
operator~$\ell_{\mathcal{E}}$ to the covering equation (see above). Let
now~$S$ be a $\tau$-shadow, where $\tau = \tau_\omega$ in
Diagram~\eqref{eq:6}. Denote by~$p_\sigma$ the fiber coordinates in the left
copy of~$\mathcal{T}\mathcal{E}$ and by~$\bar{p}_\sigma$ the same coordinates
in the right one and set $\bar{p}_\sigma^j = \tilde{D}_\sigma(S^j)$,
where~$\tilde{D}_\sigma$ is the composition of total derivatives in~$W$. This
gives the covering $\tau' = \tau_S$. The desired recursion operator is
obtained, when we set
\begin{equation*}
  S^j = \sum_{l,\sigma} S_l^{j,\sigma}p_\sigma^l + \sum_\sigma S_\sigma^j w^\sigma,
\end{equation*}
where $S_l^{j,\sigma}$, $S_\sigma^j$ are smooth functions on~$\mathcal{E}$.

\section{The main result}\label{sec:main-result}

We now pass to the equation~$\mathcal{E}$ given by~\eqref{eq:15} and choose the
functions~$x$, $y$, $t$ and
\begin{equation*}
  u_{x_i} = u_{\underbrace{x\dots x}_{i \text{ times}}},\ u_{x_i,y_j} =
  u_{\underbrace{x\dots x}_{i \text{ times}}\underbrace{y\dots y}_{j \text{
        times}}},\ u_{x_i,y_j} = 
  u_{\underbrace{x\dots x}_{i \text{ times}}\underbrace{t\dots t}_{j \text{
        times}}}, \quad i\geq 0,\ j>0,
\end{equation*}
and similar for~$v$. Then the total derivative on~$\mathcal{E}$ acquire the
form
\begin{align*}
  D_x&= \frac{\partial}{\partial x} +
       \sum_{i,j}\left(u_{x_{i+1}}\frac{\partial}{\partial u_{x_i}} +
       u_{x_{i+1},y_j}\frac{\partial}{\partial u_{x_i,y_j}} +
       u_{x_{i+1},t_j}\frac{\partial}{\partial u_{x_i,t_j}}\right)\\
  &+\sum_{i,j}\left(v_{x_{i+1}}\frac{\partial}{\partial v_{x_i}} +
       v_{x_{i+1},y_j}\frac{\partial}{\partial v_{x_i,y_j}} +
       v_{x_{i+1},t_j}\frac{\partial}{\partial v_{x_i,t_j}}\right),\\
  D_y&= \frac{\partial}{\partial y} +
       \sum_{i,j}\left(u_{x_i,y}\frac{\partial}{\partial u_{x_i}} + 
       u_{x_i,y_{j+1}}\frac{\partial}{\partial u_{x_i,y_j}} +
       D_x^iD_t^{j-1}(U)\frac{\partial}{\partial u_{x_i,t_j}}\right)\\
  &+\sum_{i,j}\left(v_{x_i,y}\frac{\partial}{\partial v_{x_i}} +
       v_{x_i,y_{j+1}}\frac{\partial}{\partial v_{x_i,y_j}} +
       D_x^iD_t^{j-1}(V)\frac{\partial}{\partial v_{x_i,t_j}}\right),\\
  D_t&= \frac{\partial}{\partial t} +
       \sum_{i,j}\left(u_{x_i,t}\frac{\partial}{\partial u_{x_i}} + 
       D_x^iD_y^{j-1}(U)\frac{\partial}{\partial u_{x_i,y_j}} +
       u_{x_i,t_{j+1}}\frac{\partial}{\partial u_{x_i,t_j}}\right)\\
  &+\sum_{i,j}\left(v_{x_i,t}\frac{\partial}{\partial v_{x_i}} +
       D_x^iD_y^{j-1}(V)\frac{\partial}{\partial v_{x_i,y_j}} +
       v_{x_i,t_{j+1}}\frac{\partial}{\partial v_{x_i,t_j}}\right),
\end{align*}
where~$V = 2 (v_y u_{xx} - v_x u_{xy})+ u_x v_{xy} - u_y v_{xx}$
and~$U = u_xu_{xy} - u_yu_{xx}$ are right-hand sides of the first and second
equations in~\eqref{eq:15}, respectively.

\subsection*{Symmetries}
\label{sec:symmetries}

The defining equation~$\ell_{\mathcal{E}}(S) = 0$ for symmetries
of~\eqref{eq:15} is
\begin{equation}
  \label{eq:7}
  \begin{array}{rcl}
    D_yD_t(\psi)
    &=& 2v_yD_x^2(\phi) - 2v_xD_xD_y(\phi)
        +v_{xy}D_x(\phi) - v_{xx}D_y(\phi) \\[2pt] 

    &&\qquad- 2u_{xy}D_x(\psi) - u_yD_x^2(\psi) + u_xD_xD_y(\psi) +
        2u_xD_y(\psi),\\[3pt]  
    D_yD_t(\phi)
    &=& u_{xy}D_x(\phi) - u_yD_x^2(\phi) + u_xD_xD_y(\phi) - u_{xx}D_y(\phi).
  \end{array}
\end{equation}
Solving System~\eqref{eq:7} for~$\phi$ and~$\psi$ of order~$\leq 3$, we get
the symmetries
\begin{align*}
  S_1 &= \big(xu_x - 2u, xv_x + 2v\big),\\
  S_2(T) &= \big(T, 0\big),\\
  S_3(T) &= \big(Tu_x + \dot{T}x, Tv_x\big),\\
  S_4(T) &= \big(\frac{1}{2}\ddot{T}x^2 + \dot{T}(xu_x  - u)+ Tu_t,
  (xv_x + 2v)\dot{T} +
        Tv_t\big),\\ 
  S_5(Y) &= \big(Yu_y, Yv_y\big),\\
  S_6 &= \big(0, u_{xxx}\big),\\
  S_7 &= \big(0,\frac{1}{2}u_{xx}^2 + u_xu_{xxx} - u_{xxt}\big),\\
  S_8 &= \big(0, u_xu_{xx}^2- u_{xt}u_{xx}+ u_x^2u_{xxx} - 2u_xu_{xxt} +
        u_{xtt}\big),\\ 
  S_9 &= \big(0,  \frac{3}{2}u_x^2u_{xx}^2 - 3u_xu_{xt}u_{xx} +
        \frac{3}{2}u_{xt}^2 -3u_x^2u_{xxt} + 3u_xu_{xtt} + u_x^3u_{xxx} -
        u_{ttt}\big),\\ 
  S_{10} &= \big(0, \frac{u_{xy}^2 - 2u_yu_{xxy}}{u_y^2}\big),\\
  S_{11} &= \big(0, \frac{u_{xy}u_{yy} - u_yu_{xyy}}{u_y^3}\big),\\
  S_{12} &= \big(0,  v\big),\\
  S_{13}(T) &= \big(0, Tu_x - \frac{1}{2}\dot{T}x\big),\\
  S_{14}(T) &= \big(0, \frac{1}{6}\ddot{T}x^2 + Tu_x^2 + (T - \frac{2}{3}\dot{T}x
           )u_x - 
  \frac{1}{6}(2u + 3x)\dot{T} - \frac{2}{3}Tu_t\big),\\
  S_{15}(T) &= \big(0, T\big),\\
  S_{16}(Y) &= \big(0, \frac{Y}{u_y^2}\big).
\end{align*}
where~$Y = Y(y)$ and~$T = T(t)$ are arbitrary smooth functions and `dot'
denotes the $t$-derivative. These symmetries commute as follows\footnote{We
  omit zero commutators. `Prime' denotes the $y$-derivative.}
\begin{align*}
  [S_1,S_2(T)]
  &= 2S_2(T), \quad
    [S_1,S_3(T)]= S_3(T),\\
  [S_1,S_6]
  &= -S_6,\quad
    [S_1,S_7]= -2S_7,\quad
    [S_1,S_8]= -3S_8,\quad
    [S_1,S_9]= -4S_9,\\
  [S_1,S_{11}]
  &= S_{11},\quad
    [S_1,S_{13}(T)]
    = -3S_{13}(T),\quad
    [S_1,S_{14}(T)]= 4S_{14}(T) - S_{13}(T),\\
  [S_1,S_{15}(T)]
  &=  - 2S_{15}(T),\quad
    [S_1,S_{16}(Y)]= 2S_{16}(Y);\\
  [S_2(T_1),S_4(T_2)
  &= -    S_2(\dot{T}_1T_2 - T_1\dot{T}_2),\quad
    [S_2(T),S_9]= -S_{15}(\dddot{T}),\\
      [S_2(T_1),S_{14}(T_2)]&= -\frac{1}{3}S_{15}(T_1\dot{T}_2 +
    2\dot{T}_1T_2);\\
  [S_3(T_1),S_3(T_2)]
  &= S_2(\dot{T}_1-T_1\dot{T}_2) - T_1\dot{T}_2),\\
  [S_3(T_1),S_4(T_2)]
  &= S_3(\dot{T}_1T_2 - \dot{T}_2T_1),\\
  [S_3(T),S_8]&= S_{15}(\dddot{T}),\quad
                [S_3(T),S_9]= 2S_{13}(\dddot{T}),\\
  [S_3(T_1),S_{13}(T_2)]&= S_{15}(\dot{T}_1T_2 + \frac{1}{2}T_1\dot{T}_2),\\
  [S_3(T_1),S_{14}(T_2)]&= \frac{2}{3}S_{13}(2\dot{T}_1T_2 +
                          T_1\dot{T_2}) - S_{15}(\dot{T}_1T_2 +
                          \frac{1}{2}T_1\dot{T}_2);\\ 
  [S_4(T_1),S_4(T_2)]&= S_4(\dot{T}_1T_2-T_1\dot{T}_2),\quad
                       [S_4(T),S_7]= -S_{15}(\dddot{T}),\\
  [S_4(T),S_8]&= -2S_{13}(\dddot{T}),\quad
                [S_4(T), S_9]= 3S_{13}(\dddot{T}) + 3S_{14}(\dddot{T}),\\
  [S_4(T_1),S_{13}(T_2)]&= -S_{13}(2\dot{T_1}T_2 + T_1\dot{T}_2),\quad
                          [S_4(T_1),S_{14}(T_2)]= -S_{14}(2\dot{T}_1T_2 + T1\dot{T}_2),\\
  [S_4(T_1),S_{15}(T_2)]&= S_{15}(2\dot{T}_1T_2 + T_1\dot{T}_2);\\
  [S_5(Y_1),S_5(Y_2)]&= S_5(Y_1'Y_2 - Y_2'Y_1),\quad
                       [S_5(Y_1),S_{16}(Y_2)]= -S_{16}(Y_2'Y_1 + 2Y_1'Y_2);\\
  [S_6,S_{12}]&= S_6;\quad
                [S_7,S_{12}]= S_7;\quad
                [S_8,S_{12}]= S_8;\quad
                [S_9,S_{12}]= S_9;\\
  [S_{10},S_{12}]&= S_{10};\quad
                   [S_{11},S_{12}]= S_{11};\quad
                   [S_{12},S_{13}(T)]= -S_{13}(T),\\
  [S_{12},S_{15}(T)]&= -S_{15}(T),\quad
                      [S_{12}S_{16}(Y)]= -S_{16}(Y).
\end{align*}

\subsection*{Tangent equation}
\label{sec:tangent-equation}

Due to \eqref{eq:7}, the tangent equation is obtained by adding to
System~\eqref{eq:15} the equations
\begin{equation}
  \label{eq:8}
  \begin{array}{rcl}
    q_{yt}
    &=& 2v_yp_{xx} - 2v_xp_{xy}+ v_{xy}p_x\\[2pt]  
     &&\qquad- v_{xx}p_y 

    - 2u_{xy}q_x - u_yq_{xx} + u_xq_{xy} + 2u_xq_y,\\[3pt]  
    p_{yt}
    &=& u_{xy}p_x - u_yp_{xx} + u_xp_{xy} - u_{xx}p_y.
  \end{array}
\end{equation}

\subsection*{Conservation laws}
\label{sec:conservation-laws}

There exist four conservation laws on~$\mathcal{T}\mathcal{E}$ of order~$\leq
2$ and linear with respect to the variables~$p_\sigma$, $q_\sigma$:
\begin{align*}
  \omega_1&= (X_1\,dx + T_1\,dt)\wedge\,dy,\quad \omega_2= (X_2\,dx +
            T_2\,dt)\wedge\,dy, \\
  \omega_3&= (X_3\,dx + Y_3\,dy)\wedge\,dt,\quad \omega_4= (X_4\,dx +
            Y_4\,dy)\wedge\,dt, 
\end{align*}
where
\begin{align*}
  &X_1 = -\frac{Yp_y}{u_y^2},\quad
    T_1 = \frac{Y(u_yp_x - u_xp_y)}{u_y^2}; \\
  &X_2 = Y(v_yp_y + u_yq_y),\quad
    T_2 = Y(v_yu_yp_x - (2v_xu_y -
    u_xv_y)p_y - u_y^2q_x + u_yu_xq_y); \\
  &X_3 = T(2u_xp_x - p_t),\quad
    Y_3 = T(u_yp_x + u_xp_y);\\
  &X_4 = T(v_xp_x + u_xq_x + q_t),\quad
    Y_4 = -T(v_xp_y + 2v_yp_x + u_yq_x - 2u_xq_y).
\end{align*}
Setting $Y = T = 1$ and after slight relabeling, we obtain the following
nonlocal variables associated with the above conservation laws:
\begin{align}\label{eq:9}
  &\begin{array}{rcl}
     w_{1,x} &=& \dfrac{p_y}{u_y^2},\\
     w_{1,t} &=& \dfrac{u_xp_y - u_yp_x}{u_y^2};\\[8pt]
     w_{2,x} &=& v_yp_y + u_yq_y,\\
     w_{2,t} &=& v_yu_yp_x + (u_xv_y- 2v_xu_y)p_y
                 u_y^2q_x + u_yu_xq_y;
   \end{array}\\\label{eq:14}
  &\begin{array}{rcl}
     w_{3,x} &=& 2u_xp_x - p_t,\\
     w_{3,y} &=& u_yp_x + u_xp_y; \\[8pt]
     w_{4,x} &=& v_xp_x + u_xq_x + q_t, \\
     w_{4,y}&
              =& 2v_yp_x - v_xp_y - u_yq_x + 2u_xq_y.
   \end{array}
\end{align}
All the nonlocal variables~$w_i$ are odd of degree~$1$.

\subsection*{Shadows}
\label{sec:shadows}

Direct computations reveal two shadows of symmetries that linearly depend on
the variables~$p_\sigma$, $q_\sigma$, and~$w_{i,\sigma}$ up to second order:
\begin{align*}
  s_0:\qquad&\bar{p}_0 = p,\quad \bar{q}_0 = q;\\
  s_1:\qquad&\bar{p}_1 = u_xp - w_3,\quad \bar{q}_1 =  v_xp - 2u_xq+
              w_4;\\ 
  s_2:\qquad&\bar{p}_2 = u_yw_1,\quad \bar{q}_2 =
              v_yw_1+\frac{w_2}{u_y^2}. 
\end{align*}
The shadow~$s_0$ is responsible for the identical operator and thus is of no
interest, while the other two ones provide nontrivial results.

\subsection*{Recursion operators}
\label{sec:recursion-operators}
Using the last formulas, we obtain the following expressions for the nonlocal
variables~$w_i$:
\begin{align}\label{eq:12}
  w_1 &= \frac{\bar{p}_2}{u_y},\quad
  w_2 = u_y(u_y\bar{q}_2  - v_y\bar{p}_2);\\\label{eq:13}
  w_3 &= u_xp - \bar{p}_1,\quad
  w_4 = -v_xp  + 2u_xq + \bar{q}_1.
 \end{align}
 Let us now substitute these expressions to the defining equations of the
 coverings~\eqref{eq:9},~\eqref{eq:14}. As the result, we obtain two
 B\"{a}cklund auto-transformations of~$\mathcal{T}\mathcal{E}$:
\begin{equation}
  \label{eq:10}
  \begin{array}{rcl}
    \bar{p}_{1,x}
    &=& u_{xx}p - u_xp_x + p_t,\\[3pt]
    \bar{p}_{1,y}
    &=& u_{xy}p - u_yp_x,\\[3pt]
    \bar{q}_{1,x}
    &=& 2v_xp_x - u_xq_x - 2u_{xx}q + v_{xx}p +
        q_t,\\[3pt]
    \bar{q}_{1,y}
    &=& 2v_yp_x - u_yq_x - 2u_{xy}q + v_{xy}p;
  \end{array}
\end{equation}
and
\begin{equation}
  \label{eq:11}
  \begin{array}{rcl}
    \bar{p}_{2,x}
    &=& \dfrac{1}{u_y}(u_{xy}\bar{p}_2 + p_y),\\[8pt]
    \bar{p}_{2,t}
    &=& \dfrac{1}{u_y}((u_xu_{xy} - u_yu_{xx})\bar{p}_2 - u_yp_x
        + u_xp_y),\\ 
    \bar{q}_{2,x}
    &=& \dfrac{1}{u_y^2}((2v_yu_{xy} + u_yv_{xy})\bar{p}_2
        + 2v_yp_y)  - \dfrac{1}{u_y}(2u_{xy}\bar{q}_2 -
        q_y),\\
    \bar{q}_{2,t}
    &=&\Big(\dfrac{u_xv_{xy}-2v_xu_{xy}}{u_y} +
        \dfrac{2u_xv_yu_{xy}}{u_y^2} -
        v_{xx}\Big)\bar{p}_2 \\
    &+&\Big(\dfrac{2v_yu_x}{u_y^2}-\dfrac{2v_x}{u_y}\Big)p_y +
              \dfrac{2(u_yu_{xx}-u_xu_{xy})}{u_y}\bar{q}_2 +
        \dfrac{u_x}{u_y}q_y - q_x. 
  \end{array}
\end{equation}
Thus, relations~\eqref{eq:10} and~\eqref{eq:11} define recursion operators for
symmetries of~Equation~\eqref{eq:15}. The $p$-components of these operators
give recursion operators for the rdDym equation
(cf.~\cite{Baran-etal-compar-stud}).

\begin{remark}\label{sec:rem-recursion-operators-1}
  Though formulas~\eqref{eq:10} and~\eqref{eq:11} look different, essentially
  they present the same object. Namely, if we resolve~\eqref{eq:11} with
  respect to the variables~$p_x$, $p_y$, $q_x$, and~$q_y$, we shall obtain
  relations~\eqref{eq:10} up to some relabeling. Consequently, the above
  presented B\"{a}cklund transformations are mutually inverse.
\end{remark}

\begin{remark}\label{sec:rem-recursion-operators-2}
  Note also that~\eqref{eq:10}, understood as a covering
  over~$\mathcal{T}\mathcal{E}$ with the nonlocal variables~$\bar{p}$
  and~$\bar{q}$ by the very construction is equivalent to the one with the
  nonlocal variables~$w_3$ and $w_4$. Explicitly, the corresponding gauge
  transformation is given by formulas~\eqref{eq:13}. In a similar
  way,~\eqref{eq:12} delivers equivalence between~\eqref{eq:11} and the
  covering with the variables~$w_1$ and~$w_2$.
\end{remark}

\subsection*{Actions}
\label{sec:actions}

Let us indicate how the operator~\eqref{eq:10} acts on symmetries of the
equation~$\mathcal{E}$; due to Remark~\ref{sec:rem-recursion-operators-1}, the
action of~\eqref{eq:11} is the opposite. Note also that of~\eqref{eq:10} is
defined up to the image of~$0$, which is~$S_2(T)$. Keeping in mind these
remarks, we have:
\begin{align*}
  &0\mapsto S_2(T)\mapsto S_3(T)\mapsto S_4(T)\mapsto\star\\
  &S_1\mapsto \star\\
  &S_5(Y)\mapsto 0\\
  &S_{11}\mapsto -\frac{1}{2}S_{10}\mapsto 2S_6\mapsto \star\\
  &S_7\mapsto -S_8\mapsto \star\\
  &S_9\mapsto \star\\
  &S_{12}\mapsto \star\\
  &S_{15}(T)\mapsto -2S_{13}(T)\mapsto \frac{3}{2}\big(S_{13}(T) -
    S_{14}(T)\big)\\ 
  &S_{14}(T)\mapsto \star\\
  &S_{16}(Y)\mapsto 0.
\end{align*}
Here~$\star$ denotes a nonlocal result.

\subsection*{Lifts of shadows}
\label{sec:lifts-shadows}

Note finally that any nonlocal symmetry in the coverings~\eqref{eq:9}
and~\eqref{eq:14} is determined by the coefficients at~$\partial/\partial u$,
$\partial/\partial v$, $\partial/\partial p$, $\partial/\partial q$,
$\partial/\partial w_1$, $\partial/\partial w_2$ (in the case
of~\eqref{eq:9}), and$~\partial/\partial w_3$, $\partial/\partial w_4$
(for~\eqref{eq:14}). Denote these coefficients by
\begin{equation*}
  U,\quad V,\quad P,\quad Q,\quad W_1,\quad W_2,\quad W_3,\quad W_4,
\end{equation*}
respectively. We state that there exist nonlocal symmetries that are the lifts
of the shadows~$s_1$ and~$s_2$ to the corresponding coverings. Namely, these
symmetries are
\begin{align*}
  \sigma_1\colon\quad
  &U= u_yw_1,\\
  &V= \frac{w_2}{u_y^2} + v_yw_1,\\
  &P= w_1p_y,\\
  &Q= w_1q_y + 2\frac{p_yw_2}{u_y^3},\\
  &W_1= w_1w_{1,y},\\
  &W_2= 2w_{1,y}w_2 - w_{2,y}w_1
    \intertext{and}
    \sigma_2\colon\quad
  &U= u_xp - w_3,\\
  &V= -2u_xq + v_xp - w_4,\\
  &P= p_xp,\\
  &Q= 2qp_x + q_xp,\\
  &W_3= pq_t + 2p_tq - v_xpp_x + u_xpq_x + 2u_xp_xq,\\
  &W_4= pp_t - 2u_xpp_x.
\end{align*}
Moreover, $\sigma_i$ are odd vector fields of degree~$1$. A direct computation
shows that the super-commutators~$[\sigma_1, \sigma_1]$
and~$[\sigma_2, \sigma_2]$ of these fields vanish, which means that the
constructed recursion operators are hereditary,~\cite{Kra-Ver-Vit-Springer}.

\section{Conclusions}
\label{sec:discussion}

The method used here to construct recursion operators for symmetries of
Equation~\eqref{eq:15} seems to be of a universal nature. In particular, it
allows to test rigorously the hereditary property of nonlocal recursion
operators. It would be interesting to apply it to other two-component
extensions of linearly degenerate equations constructed
in~\cite{Baran-etal-Lagr-deform}, as well as to the Dunajski
equation~\cite{Dun-2004}, etc.

It would be also interesting to describe (similar to how it was done
in~\cite{Baran-etal-compar-stud}) the algebra of nonlocal symmetries (and the
corresponding coverings) under the action of recursion operators described
above.

\section*{Acknowledgments}
\label{sec:acknowledgements}

Computations were done using the \textsc{Jets}~\cite{Jets} and
\textsc{Cadabra}~\cite{Cadabra, Peters-2, Peters-1} software. The authors are
grateful to O.I.~Morozov for discussion.


\begin{thebibliography}{99}
  
\bibitem{BKMV-2016} H.~Baran, I.S. Krasil{\cprime}schik, O.I.~Morozov,
  P. Voj\v{c}\'{a}k, \emph{Coverings over Lax integrable equations and their
    nonlocal symmetries}, \url{arXiv:1507.00897}, Theor.\
  Math. Phys. \textbf{188} (3) 1273--1295 (2016). Russian version \textbf{188}
  (3) 361--385 (2016), \url{https://doi.org/10.1134/S0040577916090014},
  \url{arXiv:1507.00897}.

\bibitem{Baran-etal-Lagr-deform} H.~Baran, I.S.~Krasil{\cprime}schik,
  O.I.~Morozov, P. Voj\v{c}\'{a}k, \emph{Higher symmetries of cotangent
    coverings for Lax-integrable multi-dimensional partial differential
    equations and Lagrangian deformations}, Journal of Physics: Conference
  Series, \textbf{482}, 012002, Physics and Mathematics of Nonlinear Phenomena
  2013 (PMNP2013) 22--29 June 2013, Gallipoli, Italy,
  \url{arXiv:1309.7435}
  
\bibitem{Baran-etal-compar-stud} H.~Baran, I.S.~Krasil{\cprime}schik,
  O.I.~Morozov, P. Voj\v{c}\'{a}k, \emph{Nonlocal Symmetries of Integrable
    Linearly Degenerate Equations: A Comparative Study}, Theor.\ Math.\
  Phys. \textbf{196} (2018)
  1089--1110. \url{https://doi.org/10.1134/S0040577918080019},
  \url{arXiv:1611.04938}.

\bibitem{Jets} H.~Baran, M.~Marvan, \emph{Jets. A software for differential
    calculus on jet spaces and diffieties},
  \url{https://doi.org/jets.math.slu.cz}, \url{http://jets.math.slu.cz}.

\bibitem{AMS} A.V.~Bocharov et al., \emph{Symmetries of Differential Equations
    in Mathematical Physics and Natural Sciences}, edited by A.M.~Vinogradov
  and I.S.~Krasil{\cprime}shchik). Factorial Publ.\ House, 2nd edition, 2005 (in
  Russian). English translation: Amer.\ Math.\ Soc., 1999.
    
\bibitem{Cadabra} Cadabra: a field-theory motivated approach to computer
  algebra, \url{https://cadabra.science/}
  
\bibitem{Dun-2004} M. Dunajski, \emph{A class of Einstein-Weyl spaces
    associated to an integrable system of hydrodynamic type}, J.\ Geom.\
  Phys. \textbf{51} (2004) 1, 126--137,
  \url{https://doi.org/10.1016/j.geomphys.2004.01.004},
  \url{arXiv:nlin/0311024}.

\bibitem{Fer-Mos} E.V.~Ferapontov, J.~Moss, \emph{Linearly degenerate partial
    differential equations and quadratic line complexes}, Communications in
  Analysis and Geometry, \textbf{23} (2015) no.~1, 91--127,
  \url{https://dx.doi.org/10.4310/CAG.2015.v23.n1.a3},
  \url{arXiv:1204.2777}.

\bibitem{Twente-Memo} I.S.~Krasil{\cprime}shchik, P.H.M.~Kersten P.H.M.,
  \emph{Deformations and recursion operators for evolution equations},
  Memorandum of the Twente University (1992), no. 1104, Enschede, 47 pp. Also
  in: Prastaro, A. and Rassias, Th.M. (eds.), Geometry in partial differential
  equations, World Scientific, Singapore, 1994.

\bibitem{K-V-Gdeq} I.S.~Krasil{\cprime}shchik, A.~Verbovetsky, \emph{Geometry
    of jet spaces and integrable systems}, J.~Geom. and Phys. \textbf{61}
  (2011) 9, 1633--1674 \url{arXiv:1002.0077}.

\bibitem{Ker-Kras-Kluwer} P.H.M.~Kersten, I.S.~Krasil{\cprime}shchik,
  \emph{Symmetries and recursion operators for classical and supersymmetric
    differential equations}, Kluwer Acad. Publ., Dordrecht, 2000.

\bibitem{ell-star} P. Kersten, I.S.~Krasil{\cprime}shchik, and
  A.M.~Verbovetsky, \emph{Hamiltonian operators and $\ell ^{*}$-coverings},
  J.\ Geom.\ Phys. \textbf{50} (2004), 273--302,
  \url{https://doi.org/10.1016/j.geomphys.2003.09.010},
  \url{arXiv:math/0304245}

\bibitem{Kra-Ver-Vit-Springer} I.S.~Krasil{\cprime}shchik, A.M.~Verbovetsky,
  R.~Vitolo, \emph{The Symbolic Computation of Integrability Structures for
    Partial Differential Equations}, Texts \& Monographs in Symbolic
  Computation, Springer, 2017.

\bibitem{VinKrasTrends} I.S.~Krasil{\cprime}shchik, A.M.~Vinogradov,
  \emph{Nonlocal trends in the geometry of differential equations: symmetries,
    conservation laws, and B\"{a}cklund transformations}, Acta Appl.\ Math.,
  \textbf{15} (1989) 1-2, 161--209, \url{https://doi.org/10.1007/BF00131935}.

  
\bibitem{Kuper-dark-eqs} B.A.~Kupershmidt, \emph{Dark equations}, J.\ Nonlin.\
  Math.\ Phys. \textbf{8} (2001) 3, 363--445,
  \url{arXiv:nlin/0107076}.

\bibitem{MA-Sh} L.~Mart\'{\i}nez Alonso, A.B.~Shabat, \emph{Hydrodynamic
    reductions and solutions of the universal hierarchy}, Theor.\ Math.\
  Phys. \textbf{140} (2004) 2, 1073--1085,
  \url{https://doi.org/10.1023/B:TAMP.0000036538.41884.57},
  \url{arXiv:nlin/0312043}.
  
\bibitem{Mar-another} M.~Marvan, \emph{Another look on recursion operators},
  in: Differential Geometry and Applications, Proc.\ Conf.\ Brno, 1995
  (Masaryk University, Brno, 1996) 393--402.
  
\bibitem{Mar-ort} M.~Marvan, \emph{Sufficient set of integrability conditions
    of an orthonomic system}, Found.\ Comput.\ Math. \textbf{9} (2009) 6,
  651--674.

\bibitem{Morozov2012} O.I.~Morozov, \emph{Recursion operators and nonlocal
    symmetries for integrable rmdKP and rdDym equations},
  \url{arXiv:1202.2308}.

\bibitem{Morozov2012_SIGMA} O.I.~Morozov, \emph{A two-component generalization
    of the integrable rdDym equation}, SIGMA \textbf{8} (2012), 051, 5 pp,
  \url{https://doi.org/10.3842/SIGMA.2012.051}, \url{arXiv:1205.1149}.

\bibitem{Morozov2014_CEJM} O.I.~Morozov, \emph{A recursion operator for the
    universal hierarchy equation via Cartan's method of equivalence}, Cent.\
  Eur.\ J.\ Math. \textbf{12} (2014) 2, 271--283,
  \url{https://doi.org/10.2478/s11533-013-0345-2}, \url{arXiv:1205.5748}.

\bibitem{Ovsienko2010} V.~Ovsienko, \emph{Bi-Hamiltonian nature of the
    equation} $u_{tx}= u_{xy} \,u_y - u_{yy}\, u_x$. Adv.\ Pure Appl.\ Math.,
  \textbf{1} (2010), 7--17, \url{https://doi.org/10.1515/apam.2010.002},
  \url{arXiv:0802.1818}.
  

\bibitem{Peters-2} Peeters K., Cadabra2: computer algebra for field theory
  revisited, J.\ Open Source Softw. \textbf{3} (2018) 32, 1118,
  \url{https://joss.theoj.org/papers/10.21105/joss.01118}
  
\bibitem{Peters-1} Peeters K., Introducing Cadabra: a symbolic computer
  algebra system for field theory problems, \url{arXiv:hep-th/0701238}.

\bibitem{Sergyeyev2015} A.~Sergyeyev, \emph{A simple construction of recursion
    operators for multidimensional dispersionless integrable systems}, J.\
  Math.\ Anal. Appl. \textbf{454} (2017) 2, 468--480.
  \url{arXiv:1501.01955}.

\end{thebibliography}
\end{document}